\documentclass[prb,twocolumn,showpacs,superscriptaddress]{revtex4}
\usepackage{graphicx}
\usepackage{mathrsfs}
\usepackage{amsmath,amssymb}

\begin{document}
 \title{Evanescent states in quantum wires with Rashba spin-orbit coupling}
\author{Lloren\c{c} Serra}
\affiliation{Departament de F\'{i}sica, Universitat de les Illes Balears,
  E-07122 Palma de Mallorca, Spain}
\affiliation{Institut Mediterrani d'Estudis Avan\c{c}ats IMEDEA (CSIC-UIB), 
E-07122 Palma de Mallorca, Spain}
\author{David S\'anchez}
\author{Rosa L\'opez}
\affiliation{Departament de F\'{i}sica, Universitat de les Illes Balears,
  E-07122 Palma de Mallorca, Spain}
\date{\today}
\begin{abstract}
We discuss the calculation of evanescent states in quasi-one-dimensional
quantum wires
in the presence of Rashba spin-orbit interaction. We suggest a computational
algorithm devised for cases in which longitudinal and transverse 
motions are coupled.
The dispersion relations are given for some selected cases, illustrating 
the feasibility of the proposed computational method. As a practical application,
we discuss the solutions for a wire containing a potential step.
\end{abstract}


\maketitle

\section{Introduction}

Quantum wires formed in two-dimensional electron gases are electron waveguides
allowing transmission only along one direction. Due to the lateral confinement, at 
typical electron densities a number of energy subbands may be occupied, which 
are referred to as transverse modes or channels.
Depending on the behavior of 
the electronic states along the wire
they are usually classified in propagating and evanescent. The latter
decay with the distance and, therefore, are irrelevant in the asymptotic 
regions of the quantum wire, where only propagating
modes can exist. Nevertheless, evanescent modes are of paramount importance
for a wire with inhomogeneities such as potential barriers or
wells.\cite{Bag90} This is because the evanescent modes strongly influence the
scattering conditions determining the reflection and transmission 
coefficients from a given potential inhomogeneity. 
This way, the evanescent 
modes determine the amplitudes of the propagating modes in the asymptotic 
regions of a wire with a scattering center.

The wavenumber $k$ gives the dependence with the distance $x$ along the
wire for each specific mode as $\exp(ikx)$. While for propagating modes 
$k$ is a real number, for evanescent states it must obviously have an imaginary
part. In the most general case $k$ may also contain a real part for 
evanescent modes. When transverse and longitudinal motions are decoupled
the wavenumbers for propagating and evanescent modes can be trivially
obtained. Indeed, if
transverse motion energies are quantized by a set of
$k$-independent eigenvalues $\varepsilon_n$, with $n=1,2,\dots$,
the propagating mode wavenumbers fulfill
\begin{equation}
\label{eq1}
k=\pm\sqrt{2m(E-\varepsilon_n)}/\hbar\qquad (E\ge \varepsilon_n)\;,
\end{equation}
where $E$ is the total energy. On the contrary, evanescent modes are characterized
by
\begin{equation}
\label{eq2}
k=\pm i \sqrt{2m(\varepsilon_n-E)}/\hbar\qquad (E<\varepsilon_n)\;.
\end{equation}

The above scenario of Eqs.\ (\ref{eq1}) and (\ref{eq2}) with analytically 
known wavenumbers, purely real and purely imaginary for propagating and 
evanescent modes, respectively, breaks down when motion along longitudinal
and transverse directions are coupled. In this case, transverse quantization
is not given by a fixed set of eigenvalues $\varepsilon_n$ but
must be solved explicitly for each $E$ and $k$. This 
makes, in practice, the calculation of evanescent states a formidable 
task. The mathematical difficulty is exacerbated by the fact that
for general complex $k$'s one looses standard properties such as 
the Hermitian condition of the eigenvalue equation, required in 
most matrix diagonalization algorithms. 
As a matter of fact, given $E$ the problem can be mathematically  presented 
as a non-linear eigenvalue equation for $k$.
This situation is found,
for instance, when a magnetic field is coupled with the electron's
orbital motion \cite{Barb97} or in the specific case addressed in this 
work, when a spin-orbit coupling is present. 
Evanescent states 
with spin-orbit interaction have been recently considered in 
Refs.\ \onlinecite{UB05} and \onlinecite{LB05} 
in the context of plane-wave formalisms, adequate for interfaces 
between flat bottom potentials. However, to the best of our knowledge, 
a general method to obtain
the evanescent modes in transmission channels with arbitrary transverse
confinements is lacking in the literature.   
 
We stress that the evanescent states do not fulfill
proper boundary conditions along the channel since the fact that they vanish in 
one direction necessarily implies their divergence in the reversed one.
These modes are, therefore, not physically realizable in the entire
channel, for $x$ spanning the interval $(-\infty,+\infty)$.
Nevertheless, they are extremely important because many physical states 
do behave as evanescent in restricted domains. This {\em unphysical} 
condition of the extended
evanescent states explains why, mathematically, they are solutions of 
a non Hermitian problem, while physical states always originate 
from Hermitian operators in quantum mechanics.

In this work we shall discuss a practical algorithm to obtain the 
evanescent modes when transverse and longitudinal motions are
coupled through a spin-orbit term of Rashba type,\cite{Ras60} typical of 
two-dimensional electron nanostructures. 
Although in the specific applications
we shall assume a parabolic transverse potential,  
the suggested approach will be equally valid for 
arbitrary transverse confinements. 
Numerical results for selected
values of the Rashba coupling strength will be presented as an
illustration of the method. Interest in the Rashba spin-orbit
coupling is mostly due to the tunability of the Rashba strength
by means of electric gates which opens a possibility of
spin control in nanostructures, with the Datta and Das spin transistor
as a well known spintronic device proposal.\cite{DD90} 

This paper is organized as follows. In Sec.\ II
we present the physical system. Section III is devoted to the 
practical algorithm to compute evanescent states and Sec.\ IV shows
selected numerical results and a practical application of our approach. 
Finally, Sec.\ V draws the conclusions of the 
work.

\section{Physical system}

We consider a two dimensional electron gas lying on the $xy$ plane, with 
a parabolic confinement in the $y$ direction and perfect translational
invariance along $x$. The corresponding Hamiltonian operator ${\cal H}_0$
reads
\begin{equation}
\label{eq3}
{\cal H}_0 = \frac{p_x^2+p_y^2}{2m}+\frac{1}{2}m\omega_0^2 y^2\; .
\end{equation}
Additionally, a spin-orbit coupling of Rashba type, with coupling
strength $\alpha$, is also active 
\begin{equation}
\label{eq4}
{\cal H}_R = \frac{\alpha}{\hbar}(p_y\sigma_x-p_x\sigma_y)\; .
\end{equation}
The coupling strength $\alpha$ is assumed constant throughout the system, including 
the asymptotic regions in $x$ and $y$ directions.
The total Hamiltonian is thus ${\cal H}={\cal H}_0+{\cal H}_R$ and we
are interested in the solutions of Schr\"odinger's equation for a 
given energy $E$
\begin{equation}
\label{eq5}
({\cal H}-E) \Psi(x,y,\eta)=0\; ,
\end{equation}
where $\eta=\uparrow,\downarrow$ is labelling the spin double valued variable.
Any energy is physically acceptable since the wire spectrum will be 
continuous for the propagating modes and, moreover, since we also consider
evanescent modes even negative energies will yield solutions to Eq.\ (\ref{eq5}). 

As mentioned in the introduction, the spinorial wavefunction is assumed 
to be separable in the following form
\begin{equation}
\label{eq6}
\Psi(x,y,\eta) \equiv
\phi(y,\eta)e^{ikx}\;.
\end{equation}
It is convenient to express the spin-dependent part in terms of the 
eigenspinors of $\sigma_x$, $\chi_{x\pm}(\eta)$, as
\begin{equation}
\label{eq7}
\phi(y,\eta) \equiv
\phi_1(y)\chi_{x+}(\eta) + 
\phi_2(y)\chi_{x-}(\eta)
\;.
\end{equation}

Equation (\ref{eq5}) can be recast as a matrix equation for the 
amplitudes $\phi_{1,2}(y)$,
\begin{widetext}
\begin{equation}
\label{eq8}
\left(
\begin{array}{cc}
h_0-i\alpha\frac{d}{dy}+\frac{\hbar^2k^2}{2m}-E & -i\alpha k\\
i\alpha k & h_0+i\alpha\frac{d}{dy}+{\hbar^2k^2\over 2m}-E
\end{array}
\right)
\left(
\begin{array}{c}
\phi_1(y)\\
\phi_2(y)
\end{array}
\right)
= 0\; ,
\end{equation}
\end{widetext}
where we have defined the transverse oscillator
operator
\begin{equation}
\label{eq9}
h_0 \equiv -{\hbar^2\over 2m}{d^2\over dy^2}+
\frac{1}{2}m\omega_0^2y^2\; .
\end{equation}
Equation (\ref{eq8}) is the central equation we intend to solve 
in this work. 
Although it has the formal appearance of a
linear eigenvalue equation, with eigenvalue $E$, in our case
the energy is given and what is unknown is the wavenumber $k$. 
Notice also that for complex $k$'s the matrix in
Eq.\ (\ref{eq8}) in non Hermitian, which prevents the use of
standard matrix diagonalization routines. This invalidates
the computational strategy normally used for propagating modes
and consisting in:
a) preassign a value to $k$; b) diagonalize (\ref{eq8}); 
c) find {\em a posteriori} what $k$'s give as eigenvalues the energy 
of interest $E$.

\section{The algorithm}

An essential property allowing a strategy to solve Eq.\ (\ref{eq8})
is that the amplitudes $\phi_{1,2}(y)$ must fulfill the boundary
condition that they vanish for $y\to\pm\infty$. Of course, $\phi_{1,2}(y)$
must also be smooth functions of $y$, continuous and with continuous 
derivatives for all $y$'s. These conditions can be fulfilled 
at the same time {\em only} for some specific wavenumbers representing 
propagating (real $k$'s) and evanescent (complex $k$'s) modes.

Our algorithm is based on the use of finite differences, discretizing
the $y$ axis in $N$ uniformly distributed points in the interval 
$[y_{\it min}, y_{\it max}]$.
The derivatives can then be computed using $n$-point formulas,
\cite{Abr} i.e., 
$n$ neighboring points are required to compute the derivative at each 
grid point. We transform our problem into a linear 
system of $2N$ scalar linear equations, yielding the $2N$ unknowns corresponding 
to $\phi_1$ and $\phi_2$ on the grid. This linear system 
has a solution fulfilling the boundary conditions at $y_{\it min}$ and $y_{\it max}$
for any $k$, real or complex, but only for 
the physically acceptable wavenumbers both functions have 
continuous first derivative at an arbitrarily chosen matching point
$y_m$. Since the derivatives may be discontinuous at $y_m$ 
it is essential in the calculation of the derivatives not to invoke points 
lying on opposite sides of the matching 
border. This is accomplished with the use of non-centered formulas
for the derivatives at points near the matching border. 

\begin{figure}[t]
\centering
\includegraphics*[width=6cm]{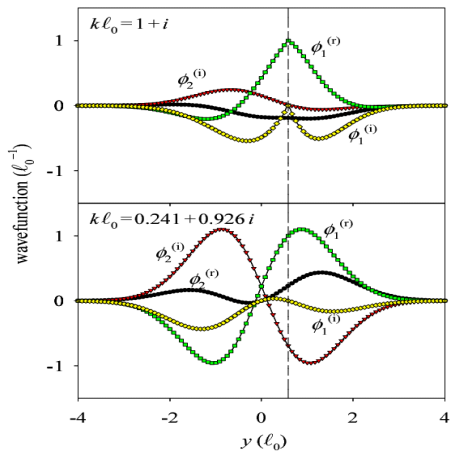}
\caption{(Color online) Wavefunction amplitudes illustrating the algorithm. Real and imaginary
parts correspond, respectively, to (r) and (i) superscripts. The total
energy is $E=\hbar\omega_0$ while the Rashba 
strength is chosen as $\alpha=0.3\hbar\omega_0\ell_0$, with $\ell_0=\sqrt{\hbar/m\omega_0}$
the oscillator length. The position of the matching point is given by the 
dashed line. Upper and lower panels correspond to the indicated wavenumbers.}
\label{fig:1}
\end{figure}

At the matching point we do not impose the discretized Eq.\ (\ref{eq8})
but, instead, we ensure a solution different from the trivial one 
(vanishing for all $y$'s) by arbitrarily choosing that $\phi_1(y_m)=1$.
An additional equation is needed to have as many equations as unknowns.
We impose continuity of the first derivative of $\phi_2$ at 
$y_m$. In summary, the resulting linear system is given by:
\begin{itemize}
\item[a)] $(2N-2)$ equations obtained by discretizing (\ref{eq8})
for $y\ne y_m$, without crossing the matching border for the derivatives,
\item[b)] $\phi_1(y_m)=1$,
\item[c)] $(d\phi_2/dy)_L-(d\phi_2/dy)_R=0$, where $(d\phi_2/dy)_{L,R}$
represent $\phi_2$ derivatives at $y_m$ using noncentered formulas with 
the left ($L$) and right ($R$) neighboring points.
\end{itemize}
The suggested algorithm does not yield normalized transverse
amplitudes but, of course, normalization
$\int{|\phi_1(y)|^2 + |\phi_2(y)|^2 dy }=1$
can be trivially imposed after the solution has been obtained.

The matrix representing the linear system is greatly sparse and can be
very efficiently solved using standard numerical routines.\cite{Harwell}
The precision of the method increases when using larger $n$'s for the 
derivatives. We have used up to 11-point formulas but, typically, 5 or 7 points
already provide quite accurate results. As an example, Fig.\ 1
shows the amplitudes $\phi_{1,2}(y)$ for two different wavenumbers: upper
panel shows a discontinuous first derivative at the matching
point, indicating that the chosen $k$ is not physically valid; the situation is
different in the lower panel, with perfectly smooth functions. 

\begin{figure}[t]
\centering
\includegraphics*[width=6cm]{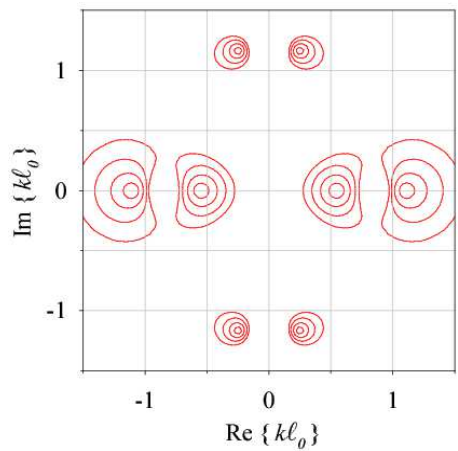}
\caption{(Color online) Contour lines of ${\cal F}(k)$ showing the position of the nodes.
We have used $E=0.75\hbar\omega_0$ and $\alpha=0.3\hbar\omega_0\ell_0$.}
\label{fig:2}
\end{figure}

Finding the physically acceptable wavenumbers is done by exploring the
complex-$k$ plane. In practice, we define the real function
\begin{equation}
\label{eq10}
{\cal F}(k)=\left\vert
 \left(\frac{d\phi_1(y_m)}{dy}\right)_L
-\left(\frac{d\phi_1(y_m)}{dy}\right)_R \right\vert\; ,
\end{equation}
and look for the zeros of ${\cal F}$ by sweeping ${\rm Re}(k)$ and ${\rm Im}(k)$,
the real and imaginary parts of the wavenumber, within a preselected range.

\section{Results}

\subsection{Symmetry considerations}

Figure 2 shows the contour lines of ${\cal F}$ circling the position of the 
nodes for selected values of $E$ and $\alpha$. Having determined the 
approximate locations, it is then a simple matter
to zoom in and accurately determine the nodes. One must be careful, however,
not to miss nodes for which the corresponding minimum is very narrow.
The figure illustrates 
a symmetry in the $k$ plane: if a mode has a certain $k$ then 
wavenumbers obtained by changing sign $\pm {\rm Re}(k)$ and/or  
$\pm {\rm Im}(k)$ are also physically valid. Similar conditions were
discussed in Ref.\ \onlinecite{Barb97} for a quantum wire in a perpendicular
magnetic field. Evanescent modes with wavenumbers  having a nonzero real
part will therefore come in groups of four, such as the modes in 
Fig.\ \ref{fig:2}  with $k=\pm0.247\pm1.166 i$, corresponding to $k$, $k^*$,
$-k$ and $-k^*$.

Mode degeneracies can be explained taking into account that if a solution 
to Eq.\ (\ref{eq8}) is characterized by a given wavenumber and amplitudes
$\{k,\phi_1,\phi_2\}$ then, by simply taking the complex conjugate of 
Eq.\ (\ref{eq8}), we find a solution having $\{k^*,\phi_2^*,\phi_1^*\}$.
On the other hand, our starting Hamiltonian ${\cal H}$ is time reversal
invariant which, for a spin 1/2 system, implies that the solutions must 
appear in degenerate pairs of time reversed states known as Kramers doublets.
Using the time reversal operator for a spin 1/2 system, $\Theta=-i\sigma_y {\cal K}$, where 
${\cal K}$ represents complex conjugation, we find the Kramers partner 
of $\{k,\phi_1,\phi_2\}$ as $\{-k^*,\phi_2^*,-\phi_1^*\}$. In summary, with
both complex conjugation (cc) and time reversal symmetry we obtain the fourfold
degeneracy
\begin{eqnarray}
\label{eq11n}
\begin{array}{c}
\{k,\phi_1,\phi_2\} \\
\Theta{\left\updownarrow\rule{0mm}{3mm}\right.}\\
\{-k^*,\phi_2^*,-\phi_1^*\}
\end{array}
&
\begin{array}{c}
\stackrel{cc}{\longleftrightarrow} \\
{\left.\rule{0mm}{3mm}\right.} \\
\stackrel{cc}{\longleftrightarrow}
\end{array}
&
\begin{array}{c}
\{k^*,\phi_2^*,\phi_1^*\} \\
-\Theta{\left\updownarrow\rule{0mm}{3mm}\right.}\\
\{-k,-\phi_1,\phi_2\}\; .
\end{array}
\end{eqnarray}

It is worth stressing that the degeneracies expressed by Eq.\ (\ref{eq11n})
are valid irrespective of the $y$-inversion symmetry of the wire potential.
This is a qualitative difference with the result for wires in perpendicular 
magnetic field,\cite{Barb97} where it was found that asymmetric wires
display asymmetric ${\rm Re}(k)$ branches, i.e., they do not
fulfill the symmetry $k\leftrightarrow -k$. The difference originates from the 
breaking of time reversal invariance by the external magnetic field.

\subsection{Mode dispersion diagrams}

\begin{figure}[t]
\centering
\includegraphics*[width=7.5cm]{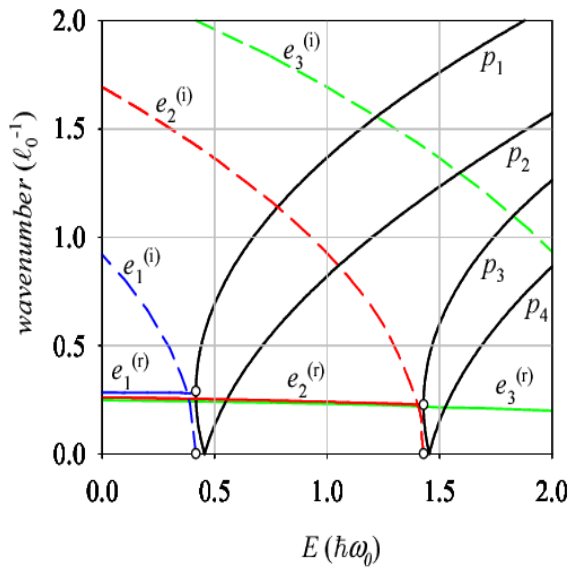}
\caption{(Color online) Mode dispersion for $\alpha=0.3\hbar\omega_0\ell_0$.
Branch labels indicate mode number $n$ as well as evanescent or 
propagating character as $e_n$ and $p_n$, respectively. In the case of
evanescent modes, superindexes $(r)$ and $(i)$ are used to indicate whether
the branch gives the real or the imaginary part of the wavenumber.}
\label{fig:3}
\end{figure}

The mode dispersion for the same wire of the preceding figures is shown in 
Fig.\ \ref{fig:3}. We only plot positive wavenumbers
noting that, with the above mentioned symmetry, wavenumber signs can be 
inverted giving a four-fold degeneracy of each evanescent mode and two-fold
for the propagating ones. The dispersion of propagating modes shows a 
familiar picture, already investigated in 
detail.\cite{MB99,MK01,GZ02,VR03,PN04,SS05} 
There is a threshold energy $E_n^{({\it th})}$ for the activation of the $n$-th 
propagating mode.
Due to the spin-orbit coupling, when $E$ slightly exceeds $E_n^{({\it th})}$, 
propagating
states with nonzero wavenumber $k_n^{({\it th})}$, belonging to the $n$-th mode, 
are allowed. 

Surprisingly, the evanescent modes in Fig.\ \ref{fig:3} resemble 
those of {\em asymmetric} wires in vertical
magnetic field discussed in Ref.\ \onlinecite{Barb97}, in spite
of the present wires being symmetric and having no field.
For $E$ slightly below $E_n^{({\it th})}$ there is an evanescent mode whose
${\rm Im}(k)$ approaches zero as $E$ approaches $E_n^{({\it th})}$
from below but, quite remarkably, ${\rm Re}(k)$ remains essentially stuck
at the threshold value for the propagating mode, $k_n^{({\it th})}$. 
This is true
even for energies much smaller than $E_n^{({\it th})}$. 
Notice that, on the contrary, for evanescent states
${\rm Im}(k)$  rapidly increases when 
$E$ decreases, 
indicating a faster decay of the evanescent mode when 
the energy separates from the threshold $E_n^{({\it th})}$.
Strictly speaking, the real part also varies with $E$ although it can
be hardly noticed to the scale of Fig.\ \ref{fig:3}.

\begin{figure}[t]
\centering
\includegraphics*[width=7.5cm]{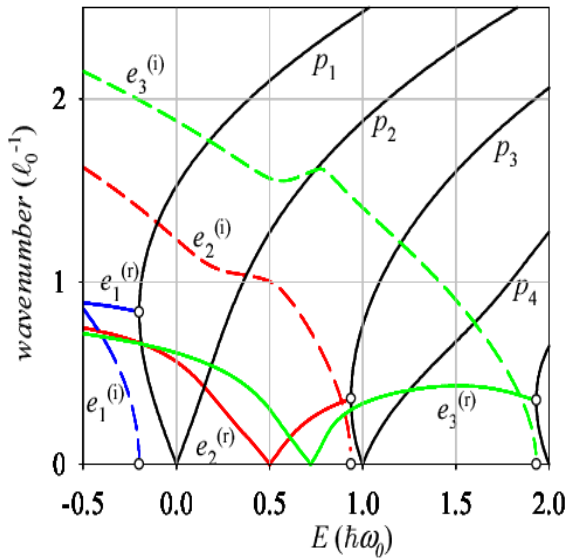}
\caption{(Color online) Same as Fig.\ \ref{fig:3} for 
$\alpha=\hbar\omega_0\ell_0$.}
\label{fig:4}
\end{figure}

Figure \ref{fig:4} shows the mode dispersion for a case of strong spin-orbit 
coupling, when $\alpha=\hbar\omega_0\ell_0$. As compared to the weak 
coupling case (Fig.\ \ref{fig:3}),
the mode branch distribution now looks more complicated. Nevertheless, it 
remains true that below the threshold of a propagating mode there is 
an evanescent mode whose wavenumber initially coincides with $k_n^{({\it th})}$, 
the wavenumber of the propagating mode at threshold. 
If the energy is further 
decreased, ${\rm Re}(k)$ does not remain almost constant as in Fig.\ \ref{fig:3},
but displays a rather strong variation, first decreasing and then 
increasing again. It is worth stressing that the symmetry discussed above 
still applies to Fig.\ \ref{fig:4} 
and one can obtain valid wavenumbers by reverting signs in either ${\rm Re}(k)$ or
${\rm Im}(k)$. Another conspicuous feature of Fig.\ \ref{fig:4}
is that $e_n^{(i)}$, the branches giving ${\rm Im}(k)$, are not monotonously increasing with decreasing energy
but display a marked wiggle at $E\sim 0.4\hbar\omega_0$ and 
$0.6\hbar\omega_0$ for $e_2^{(i)}$ and $e_3^{(i)}$, respectively. 

The simpler behavior of the weak coupling case suggests 
an analytical expression for the evanescent mode wavenumbers. We have
indeed found that when $\alpha/(\hbar\omega_0\ell_0)\ll 1$ the evanescent 
mode wavenumbers can be well approximated by
\begin{equation}
\label{eq11}
k= \pm k_n^{(\it th)} 
\pm i \sqrt{2m(E_n^{(\it th)}-E)}/\hbar \quad (E<E_n^{(\it th)}) \; ,
\end{equation}
in terms of the threshold energy $E_n^{(\it th)}$ and wavenumber $k_n^{(\it th)}$
of the $n$-th propagating mode. Noting now that in the considered
limit one has $k_n^{(\it th)}\approx k_R$ and $E_n^{(\it th)}\approx \varepsilon_n$,
where we define the {\em Rashba wavenumber} $k_R=\alpha m/\hbar^2$ and 
the transverse oscillator
energies $\varepsilon_n=(n+1/2)\hbar\omega_0$, we can further simplify
Eq.\ (\ref{eq11}) to 
\begin{equation}
\label{eq12}
k= \pm k_R \pm i \sqrt{2m(\varepsilon_n-E)}/\hbar \quad (E<\varepsilon_n)\;.
\end{equation}
This is a very appealing result, stating that in the weak coupling case 
the only modification to Eq.\ (\ref{eq2}) introduced by the Rashba coupling 
is to add a constant (energy independent) real part given by the 
Rashba wavenumber. This result agrees with the findings of Ref.\ \onlinecite{LB05},
where the density oscillations at the interface between a ferromagnet and a 
semiconductor with Rashba interaction were found to have an
energy independent wavelength $\lambda\sim k_R^{-1}$ in the limit of low Rashba coupling.

Equation (\ref{eq12}) can be understood noting that in the weak spin-orbit limit
one can neglect the Rashba intersubband coupling term, proportional to $p_y\sigma_x$.
In this approximation the Hamiltonian is diagonal in the $\sigma_y$ basis and 
the corresponding Schr\"odinger's equation for {\em any} $k$ implies 
\begin{equation}
\label{D1}
\varepsilon_n+\frac{\hbar^2 k^2}{2m}\pm \frac{\hbar^2 k_R k}{m} -E=0\; ,
\end{equation}
whose solutions for $E<\varepsilon_n$ and neglecting terms in $k_R^2$ coincide with 
Eq.\ (\ref{eq12}).

\begin{figure}[t]
\centering
\includegraphics*[width=7.5cm]{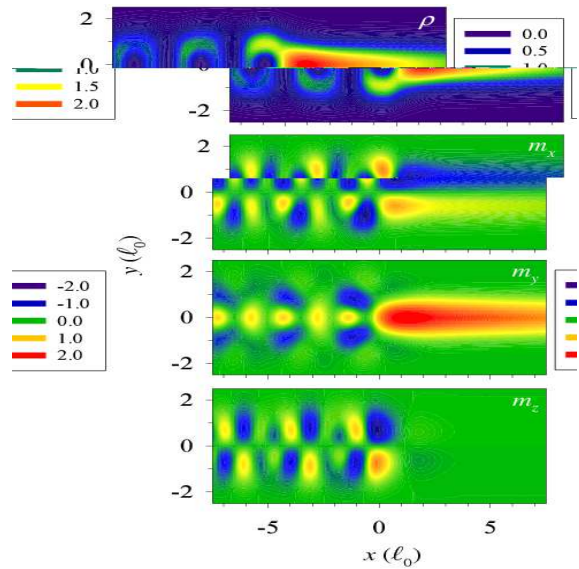}
\caption{(Color online) Density and spin magnetization distributions in a
wire with a potential step at $x=0$ of $V_0=1.13\hbar\omega_0$ and 
having $\alpha=\hbar\omega_0\ell_0$. 
The density contour numerical values are given in units of $\ell_0^{-2}$
while those of magnetization density are in units of $\hbar/(2\ell_0^2)$.
Incidence is from the left in mode $p_1$ shown in Fig.\ \ref{fig:4}, for an 
energy $E=0.93\hbar\omega_0$ slightly below the threshold for propagating modes $p_3$ and $p_4$.}
\label{fig:5}
\end{figure}

\subsection{The potential step }

To illustrate the use of the computed evanescent modes with a
specific example we have obtained the wave functions in a quantum 
wire with Rashba interaction and containing a potential step $V_0 \Theta(x)$.\cite{theta} 
Electrons incident from the left ($x<0$) impinge on the step border and, 
assuming 
their energy is not enough to allow for propagation in the right region ($x>0$), 
only pure evanescent modes will be seen for $x>0$. On the contrary, the left region 
will contain three types of modes: incident propagating, reflected propagating and 
evanescent.
The mode wavenumbers before and after the step can be obtained from
the mode dispersion diagrams, Figs.\ \ref{fig:3} and \ref{fig:4}, 
associating to each region an energy measured from the potential bottom, 
i.e., the {\em left} $E_l=E$ and {\em right} $E_r=E-V_0$ energies. 
At the step edge the wave functions must match adequately, 
this condition 
determining the amplitudes of the reflected propagating modes 
for $x<0$ and of the 
evanescent waves for all $x$. The Appendix details 
the resulting equations and their practical resolution method.
Evanescent states are crucial in this problem for
without them 
it is not possible to fulfill the matching conditions at $x=0$ and for any
value of the transverse coordinate $y$ (see Appendix). Besides, evanescent states must be considered 
in order to satisfy the obvious requirement that the reflected flux coincide with the 
incident one.

Figure \ref{fig:5} displays the density and spin magnetizations 
obtained after solving the linear system of equations corresponding to
the matching conditions.
We have assumed the same intense Rashba coupling $\alpha=\hbar\omega_0\ell_0$ of Fig.\ \ref{fig:4}, 
and a left incidence of unit flux from mode $p_1$, with energy $E=0.93\hbar\omega_0$, 
slightly below the propagation threshold for modes 
$p_3$ and $p_4$. 
The step is chosen as $V_0=1.13\hbar\omega_0\ell_0$,
which implies that the wavenumbers are those of Fig.\ \ref{fig:4}
at energies $(E_l,E_r)=(0.93,-0.20)\hbar\omega_0$.
The spatial distribution of density and magnetization
due to evanescent modes are clearly seen for $x>0$, while on the left
side there are marked interference effects between the coexisting modes.
Actually, the latter patterns are similar to those discussed in Ref.\ \onlinecite{LB05}
when considering a different physical system; namely, an
interface between a ferromagnet and a semiconductor.
Focussing on the evanescent mode side, there is a strong injection of 
spin $y$ magnetization $m_y$, clearly due to the fact that the incident
mode $p_1$ is mostly polarized along $+y$ (it is not completely polarized 
due to the admixture induced by the Rashba intersubband coupling term $\alpha p_y\sigma_x$).
There is also an important accumulation of evanescent $m_x$ magnetization,
of different signs on the two sides of the wire. Had we considered incidence
from mode $p_2$ at the same energy, the spin magnetizations would be reverted with respect to those
of Fig.\ \ref{fig:5}, indicating that the average for incidence from the two modes
$p_1$ and $p_2$ does not produce any net magnetization, not even locally. 
On the contrary, the density shown in the 
upper panel of Fig.\ \ref{fig:5} is the same for incidence from any of the two modes. 

The vanishing spin magnetization when adding 
the contributions from the two left incident channels $p_1$ and $p_2$ can be 
interpreted, theoretically, as a manifestation of the time reversal
symmetry which is conserved by the Rashba interaction. Indeed, when $E_r$ 
is such that there can be no propagation to the right all incident flux 
is reflected backwards, yielding a solution which is invariant by time reversal.

\begin{figure}[t]
\centering
\includegraphics*[width=5.5cm]{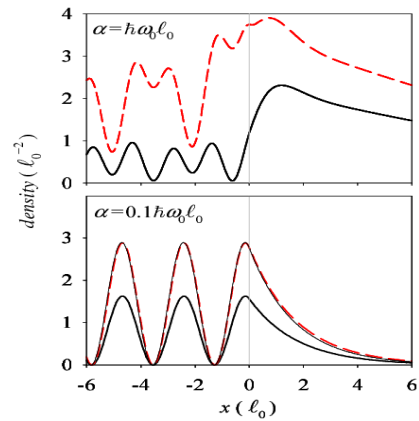}
\caption{(Color online) Density along the $y=0$ cut (thick solid) and integrated in $y$,
$\int_{-\infty}^\infty{dy \rho(y)}/\ell_0$ (dashed).
Upper panel corresponds to the solution given in the preceding figure upper panel. Lower
panel corresponds to a weaker Rashba coupling, with parameters
$\alpha=0.1\hbar\omega_0\ell_0$, $(E_l,E_r)=(1.45,0.45)\hbar\omega_0$, corresponding to
an energy 
$E=1.45\hbar\omega_0$ and step height $V_0=\hbar\omega_0$.
The thin solid line of the lower panel, almost superimposing on the dashed line, corresponds to
the strict 1D problem.}
\label{fig:6}
\end{figure}

A remarkable feature in Fig.\ \ref{fig:5} is that the density takes its maximum value 
for $x>0$, the evanescent mode side. This is totally unexpected for purely 
exponentially decaying modes.
However, since the evanescent wavenumbers with Rashba coupling are complex, we
may find a superposition of oscillating and exponentially decaying distributions, as that of
Fig.\ \ref{fig:5}. This behavior is a 
peculiarity 
of intense Rashba couplings. To further clarify this
we plot in Fig.\ \ref{fig:6} the densities for weak and strong values of $\alpha$. In the weak
$\alpha$ case the evanescent density is indeed given by a pure exponential decay that can be explained 
taking the wavenumbers given by Eq.\ (\ref{eq12}). Since ${\rm Re}(k)$ is constant, it amounts
to a common phase for all modes and thus irrelevant for the density, remaining only the pure exponentially 
decaying contributions.
The lower panel also shows the density in the strict 1D limit, when all transverse motion is neglected 
and the Hamiltonian reduces to ${\cal H}_{1D}=p_x^2/2m-\alpha p_x\sigma_y/\hbar$. The excellent agreement
with the $y$-integrated 2D density clearly proves that in the limit of small $\alpha$ the problem becomes
effectively one dimensional, as should be expected.

\begin{figure}[t]
\centering
\includegraphics*[width=5.5cm]{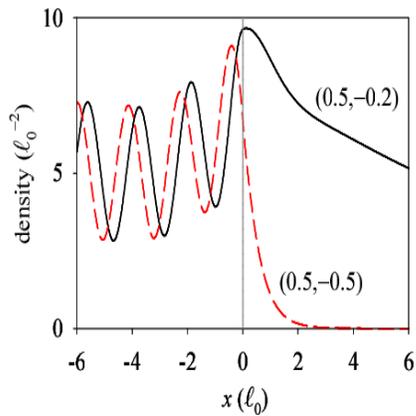}
\caption{(Color online) $y$-integrated densities 
$\int_{-\infty}^\infty{dy \rho(y)}/\ell_0$ for the shown values
of $(E_l,E_r)$ in $\hbar\omega_0$ units for the strong coupling 
limit $\alpha=\hbar\omega_0\ell_0$. Left incidence of unit flux from
modes $p_1$ and $p_2$ is assumed.}
\label{fig:7}
\end{figure}

The relevance of the evanescent modes in the strong $\alpha$ case 
manifests itself in Fig.\ \ref{fig:6} as conspicuous accumulations of 
the integrated density (dashed line), slowly decaying towards both sides 
of the edge. The decay is quite sensitive to the values of $E$ and $V_0$
or, equivalently, to the energies $(E_l,E_r)$ fixing the left and right 
modes. This is clearly seen in Fig. \ref{fig:7}, which displays the 
$y$-integrated densities for $(E_l,E_r)=(0.5,-0.2)\hbar\omega_0$
and $(0.5,-0.5)\hbar\omega_0$. Towards the left the decay is now faster
than in the upper panel of Fig.\ \ref{fig:6}, due to the higher values
of ${\rm Im}(k)$ for this $E_l$. The same argument explains
the faster decay towards the right for $E_r=-0.5\hbar\omega_0$.
In general, we find that $E_l$ and $E_r$ determine the decay rate
towards the left and right sides rather independently.

\begin{figure}[t]
\centering
\includegraphics*[width=6.5cm]{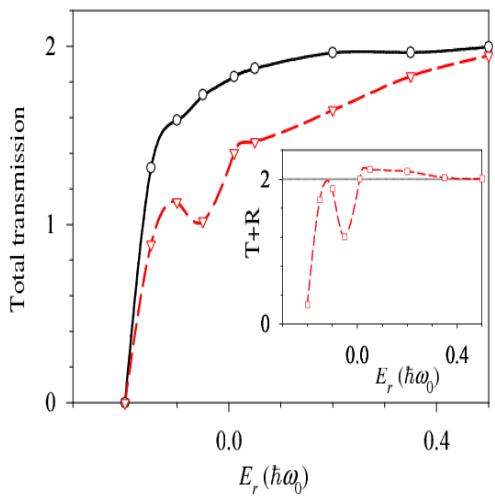}
\caption{(Color online) Transmission for left incidence in modes $p_1$ and $p_2$ on the 
potential step when 
$\alpha=\hbar\omega_0\ell_0$ and $E_l=0.93\hbar\omega_0$. The step height for each 
value of $E_r$ is $V_0=E_l-E_r$. Circles (solid line) are the complete result
while triangles (long-dashed) totally neglect the evanescent modes.
Transmission plus reflection is exactly two in the complete calculation and is 
shown in the inset when evanescent modes are neglected. The deviation
from the exact value shows the relevance of the evanescent mode contributions.}
\label{fig:8}
\end{figure}

Most interestingly, 
the evanescent modes are crucial even when the energy $E_r$ exceeds 
the propagation threshold to the right of the step. This is shown in 
Fig.\ \ref{fig:8}, where the total transmission for left
incidence in both modes $p_1$ and $p_2$ is given for the strong 
coupling case. We have fixed $E_l=0.93\hbar\omega_0\ell_0$ and increased
$E_r$ above the propagation threshold of modes $p_1$ and $p_2$.
In practice, this corresponds to lowering the step height $V_0=E_l-E_r$
while keeping fixed the incident energy $E=E_l$.
The triangles are the result
when evanescent modes are totally neglected in subsets 
$\{\tilde\phi_n,\tilde{k}_n\}$ and $\{\hat\phi_n,\hat{k}_n\}$
of the Appendix. There are sizeable differences with the transmission 
obtained when including enough evanescent states in both subsets (circles). 
Quite remarkably,
the condition of flux conservation which, for the present situation, implies that 
transmission plus reflection amount to a value of two
unit fluxes (corresponding to the incidence from modes $p_1$ and $p_2$) 
is only fulfilled by the complete calculation.
Neglecting the evanescent modes leads to a violation of this balance,
as shown by the inset in Fig.\ \ref{fig:8}. Deviations are greater
near the propagation threshold where, indeed,  it is expected that evanescent 
modes are more important.

\section{Conclusions}

Evanescent states along a quasi-one-dimensional channel couple with the
states of transverse motion when spin-orbit (Rashba) interaction is active. 
This greatly 
complicates the determination of the evanescent state wavenumbers and
wavefunctions since they 
do not obey a standard Hermitian eigenvalue problem. We have devised an algorithm
based on the requirement that the transverse spinor wavefunctions fulfill boundary 
conditions, yielding a set of linear equations that is solvable for any complex wavenumber. 
Physically acceptable wavenumbers are those having continuous derivative at an 
arbitrarily chosen matching point.

Evanescent modes in Rashba wires are characterized by complex wavenumbers, 
having real and imaginary parts. 
Due to symmetry, one can invert signs of either ${\rm Re}(k)$ or ${\rm Im}(k)$,
or both, and still obtain physically valid wavenumbers.
When decreasing the energy the wire modes evolve in
the following way: 
\begin{itemize}
\item[a)] at high energies the mode is propagating; 
\item[b)] when the energy crosses the
propagation threshold the mode becomes evanescent, with a continuous evolution of 
${\rm Re}(k)$, while ${\rm Im}(k)$ suddenly starts to grow from zero;
\item[c)] in the limit of weak Rashba coupling ${\rm Re}(k)$ for evanescent modes 
is constant and coincides with 
$k_R$, the propagation wavenumber at threshold, while ${\rm Im}(k)$ is given 
by Eq.\ (\ref{eq2}), the result of uncoupled transverse and longitudinal motions;
\item[d)] for intense Rashba couplings a complicated evolution of ${\rm Re}(k)$ and ${\rm Im}(k)$ with 
energy is observed.
\end{itemize}

A specific example illustrating the relevance of evanescent states has been solved, namely the case of 
a potential step in a wire with intense spin-orbit coupling. 
Density accumulations at the edge, as well as 
distributions of
spin magnetization have been obtained. Evanescent states are necessary to fulfill the boundary 
conditions at the step edge and maintain flux conservation. 
A remarkable feature of the nontrivial evanescent ${\rm Re}(k)$'s
is that maximal densities can be located on the evanescent mode side.
We have also analyzed the case when transport to the right is allowed,
finding an important contribution of the evanescent states to the transmissions. 

To summarize, 
the evanescent modes obtained in this work allow the investigation
of spin and density distributions around inhomogeneities and at interfaces in quantum wires with 
extended spin-orbit coupling, a condition met in many spintronic devices.

\appendix*
\section{}

In this Appendix we detail the matching conditions corresponding to the potential 
step in a quantum wire with Rashba interaction discussed in Sect.\ IV.C. 
The general wavefunction is given by a superposition of modes, each one characterized
by a transverse wavefunction and wavenumber $\{\phi_n(y,\eta),k_n\}$. 
The potential step $V_0\Theta(x)$ is such that with the given energy $E<V_0$
for left incident electrons only evanescent modes can survive in the right side.
Let us distinguish
the following three sets of states: $\{\phi_n,k_n\}$, the propagating modes incident 
from $x<0$; $\{\tilde\phi_n,\tilde{k}_n\}$, reflected propagating and evanescent modes, the 
latter vanishing for $x\to -\infty$ [with negative ${\rm Im}(k)$]; and  
$\{\hat\phi_n,\hat{k}_n\}$, evanescent modes vanishing for $x\to+\infty$. 
The most general state then reads
\begin{widetext}
\begin{equation}
\label{eq14w}
\Psi(x,y,\eta) =
\left\{
\begin{array}{lr}
\sum_n{a_n \phi_n(y,\eta) e^{i k_n x}}
+ \tilde\sum_n{b_n \tilde\phi_n(y,\eta) e^{i \tilde{k}_n x}} & \qquad{\rm if}\; x<0\; ,\\
\hat\sum_n{c_n \hat\phi_n(y,\eta) e^{i \hat{k}_n x}}& \qquad {\rm if}\; x>0\; ,
\end{array}
\right.
\end{equation}
where $a_n$, $b_n$ and $c_n$ are the usual incidence, reflection and transmission amplitudes.

The matching conditions 
at $x=0$ require continuity of the wave function and its $x$-derivative, 
\begin{eqnarray}
\label{eq15a}
\sum_n{a_n \phi_n(y,\eta)}
+
\tilde{\sum_n}{b_n \tilde\phi_n(y,\eta)} &=& 
\hat{\sum_n}{c_n \hat\phi_n(y,\eta)}\; ,\\
\label{eq15b}
\sum_n{a_n k_n \phi_n(y,\eta)}
+
\tilde{\sum_n}{b_n \tilde{k}_n \tilde\phi_n(y,\eta)} &=&   
\hat{\sum_n}{c_n \hat{k}_n \hat\phi_n(y,\eta)} \; .
\end{eqnarray}

Equations (\ref{eq15a}) and (\ref{eq15b}) must determine the reflection 
and transmission coefficients, $b_n$ and $c_n$, in terms 
of the incident ones $a_n$ and they amount to a linear system of two equations
for each value of $y$ and $\eta$.
Since we have in principle an infinite set of $y$ values, one needs also an infinite 
set of amplitudes $b_n$ and $c_n$ in order to have as many equations as unknowns. However, 
at a given energy the number of propagating states is always finite, which is illustrating
the fact that one needs to include the evanescent modes, also an infinite set, 
to fulfill Eqs.\ (\ref{eq15a}) and (\ref{eq15b}) for arbitrary $y$. 
In practice we truncate the sums over evanescent states
$\tilde\sum_n$ and $\hat\sum_n$ 
and project Eq.\ \ref{eq15a} on the set $\{\tilde\phi_n\}$
and Eq.\ \ref{eq15b} on $\{\hat\phi_n\}$. The resulting 
linear system reads
\begin{eqnarray}
\label{eq16a}
\tilde\sum_n{[\tilde\phi\tilde\phi]_{mn} b_n} -
\hat\sum_n{[\tilde\phi\hat\phi]_{mn} c_n} 
&=& - {\sum_n}{[\tilde\phi\phi]_{mn} a_n}\; ,\\
\label{eq16b}
\tilde\sum_n{\tilde{k}_n[\hat\phi\tilde\phi]_{mn} b_n} -
\hat\sum_n{\hat{k}_n[\hat\phi\hat\phi]_{mn} c_n} 
&=& - {\sum_n}{k_n[\hat\phi\phi]_{mn} a_n}\; ,
\end{eqnarray}
\end{widetext}
where we have introduced the following notation for overlap matrices
\begin{equation}
{[\tilde\phi\tilde\phi]}_{mn} = 
\sum_\eta \int{dy \tilde\phi_m^*(y,\eta) \tilde\phi_n(y,\eta)}  \; ,
\end{equation}
with obvious extensions for ${[\tilde\phi\hat\phi]}$,
${[\tilde\phi\phi]}$, ${[\hat\phi\tilde\phi]}$,
${[\hat\phi\hat\phi]}$ and 
${[\hat\phi\phi]}$.
Including enough evanescent states $\{\tilde\phi_n\}$ and $\{\hat\phi_n\}$ the 
calculation converges, giving an increasingly better fulfillment of the 
matching conditions and of the flux conservation.

The formalism can be trivially extended to consider the case when 
the energy is such that propagating modes after the step ($x>0$)
are allowed. One just needs to add the corresponding
propagating modes to the set $\{\hat\phi_n,\hat{k}_n\}$ while all 
remaining equations of the Appendix are unchanged.

\section*{ACKNOWLEDGEMENTS}

This work was supported by the Grant No.\ FIS2005-02796
(MEC) and the Spanish ``Ram\'on y Cajal'' program. 



\begin{thebibliography}{10}
\bibitem{Bag90} P. F. Bagwell, Phys.\ Rev.\ B {\bf 41}, 10354 (1990).
\bibitem{Barb97} J. C. Barbosa and P. N. Butcher,
Superlattices and Microstructures {\bf 22}, 325 (1997).
\bibitem{UB05} G. Usaj and C. A. Balserio, Europhys.\ Lett.\ {\bf 72}, 631 (2005). 
\bibitem{LB05} M. Lee and C. Bruder, Phys.\ Rev.\ B {\bf 72}, 045353 (2005). 
\bibitem{Ras60} E. I. Rashba, Fiz.\ Tverd.\ Tela.\ (Leningrad) {\bf 2}, 1224 (1960) 
[Sov.\ Phys.\ Solid State {\bf 2}, 1109 (1960)].
\bibitem{DD90} S. Datta, B. Das, Appl.\ Phys.\ Lett.\ {\bf 56}, 665 (1990).
\bibitem{Abr} E. M. Abramowitz and I. A. Stegun, eds., {\em Handbook of Mathematical Functions},
(Dover, 1972).
\bibitem{Harwell} Harwell subroutine library.
\bibitem{MB99} A. V. Moroz and C. H. W. Barnes, Phys.\ Rev.\ B {\bf 60}, 14272 (1999). 
\bibitem{MK01} F. Mireles and G. Kirczenow, Phys.\ Rev.\ B {\bf 64}, 024426 (2001).
\bibitem{GZ02}  M. Governale and U. Z\"ulicke, Phys.\ Rev.\ B {\bf 66}, 073311 (2002).
\bibitem{VR03} M. Val\'{\i}n-Rodr\'{\i}guez, A. Puente, Ll.\ Serra, Eur.\ Phys.\ J. B {\bf 34}, 359 (2003).
\bibitem{PN04} Yu.\ V. Pershin, J. A. Nesteroff, and V. Privman, Phys.\ Rev.\ B {\bf 69}, 121306(R) (2004).
\bibitem{SS05}Ll.\ Serra, D. S\'anchez and R. L\'opez, Phys.\ Rev.\ B {\bf 72}, 235309 (2005).
\bibitem{theta} We assume $\Theta(x)=0$ for $x<0$ and $\Theta(x)=1$ for $x>0$.
\end{thebibliography}
\end{document}